\documentclass[a4paper,11pt]{article}
\usepackage{pos}
\usepackage{cleveref}
\usepackage{amsmath}
\usepackage{graphicx}
\usepackage{gensymb}
\newlength{\bibitemsep}
\newlength{\bibparskip}
\let\oldthebibliography\thebibliography
\renewcommand\thebibliography[1]{%
  \oldthebibliography{#1}%
  \setlength{\parskip}{\bibitemsep}%
  \setlength{\itemsep}{\bibparskip}%
}

\title{Telescope Tilt System for the POEMMA Balloon Radio Mission}
\ShortTitle{Telescope Tilt System for PBR}
\author*[a]{Lawrence Wiencke}
\author[a]{James Brague}
\author[a]{Ash Fox-Smith}
\author[a]{Auston Froid}
\author[a]{Levi Bar-On}
\author[b]{Stephen Meyer}
\author[a]{Josh Moses}
\author[b]{Ben Stillwell}

\onbehalf{for the JEM-EUSO Collaboration}

\affiliation[a]{Colorado School of Mines, Golden, CO, USA}
\affiliation[b]{University of Chicago, KICP,
Chicago, IL, USA}

\emailAdd{lwiencke@mines.edu}

\abstract{The Probe of Extreme Multi-Messenger Astrophysics Balloon with Radio mission (PBR) will point above Earth's limb to measure PeV energy cosmic rays, and record star images to monitor optical focusing in situ. PBR will point below Earth's limb to search for earth-skimming neutrinos. PBR will also measure EeV energy cosmic rays by tilting as far down as the nadir direction. All of these searches will require changing and measuring the tilt angle of a single large integrated telescope and radio antenna assembly in the near space environment at 33 km above sea level over a mission duration as long as 50 days. In addition, the 1.1 m diameter entrance pupil of the telescope will be covered during the day by a shutter system to prevent sunlight from damaging the camera systems and opened at night to collect data. Here we present the design and status of the tilting system, the tilting monitors, the shutter system, the controller, and the preflight thermal vacuum testing process. The work draws on the experience of the 2023 Extreme Universe Space Observatory on a Super Pressure Balloon 2 mission.}

\ConferenceLogo{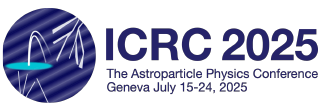}

\FullConference{%
39th International Cosmic Ray Conference (ICRC2025)\\
  15 - 24 July 2025\\
  Geneva, Switzerland}

\begin{document}

\maketitle

\section{Introduction}
The Probe of Extreme Multi-Messenger Astrophysics(POEMMA) Balloon Radio (PBR)\cite{ICRC2025Eser} is an ambitious follow-up to the Extreme Space Observatory on a Super Pressure Balloon (EUSO-SPB2)\cite{FT-paper,Adams:2025owi}, and a precursor to a space-based satellite mission such as POEMMA\cite{POEMMA:2020ykm}. PBR will measure extensive air showers (EASs) from cosmic rays through Cherenkov emission \cite{EASCherSim} and radio emission and via fluorescence emission, search for astrophysical tau neutrinos via the earth-skimming technique using optical and radio methods, and measure related backgrounds. Of particular interest are searches for neutrino emission from transient sources \cite{Reno:2019jtr,Venters:2019xwi,Venters_2021ICRC} in a "Target of Opportunity" mode. The payload is scheduled to fly from the NASA balloon launch site in W\=anaka New Zealand in 2027. The balloon will fly at 33 km for up to 50 days. Balloon flights from this NASA launch site experience dark night skies that are necessary to measure EAS optical signals above the background.

\begin{figure}[!h]
\centering
\includegraphics[width=0.9\textwidth]{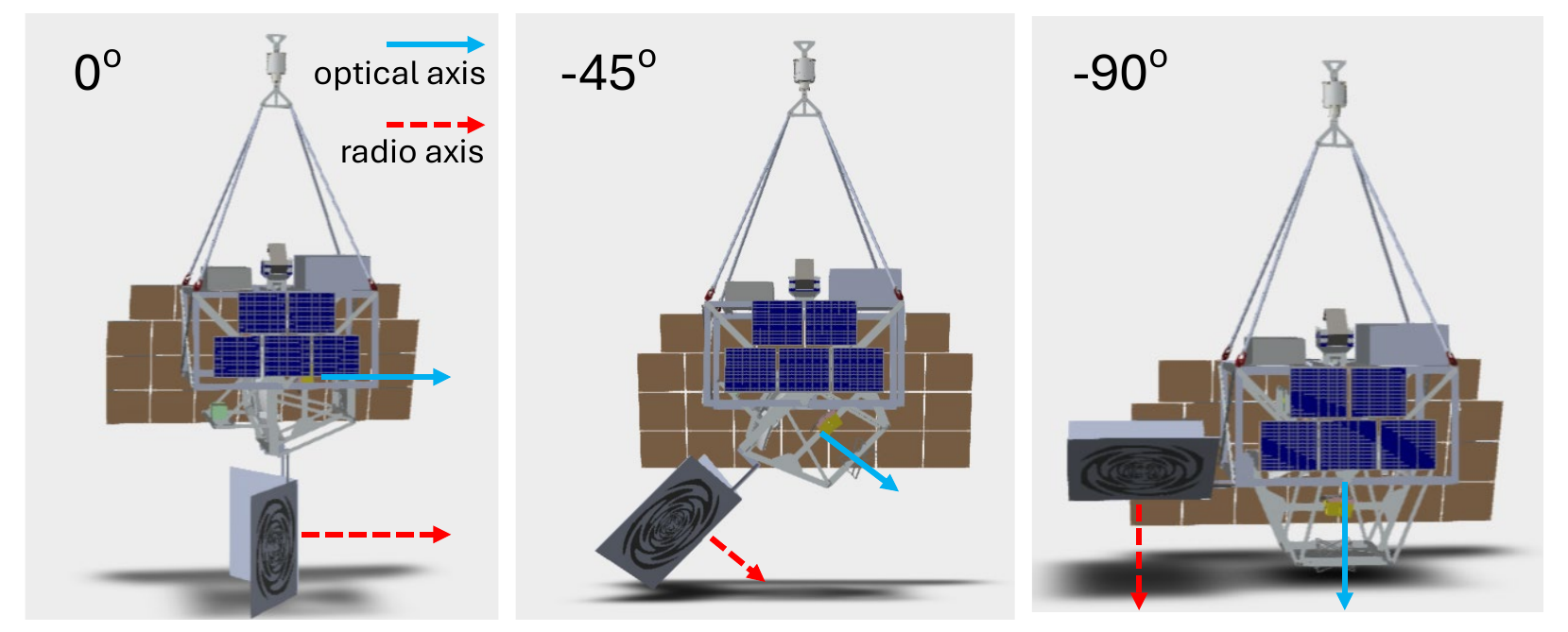}
\caption{ The POEMMA Balloon Radio payload design is shown with the telescope assembly tilted at various angles.}
\label{fig:PBR-Tilt-Angles}
\end{figure}

 Details of the PBR payload and instrumentation are available in \cite{Eric:ICRC,Francesco:ICRC, Valentina:ICRC}. Briefly, the payload (Fig. \ref{fig:PBR-Tilt-Angles}) features a modified Schmidt telescope with a 1.1~m diameter entrance pupil, a segmented spherical primary mirror, and two high-speed custom camera systems at the focal surface. The Cherenkov camera (CC) system uses an array of silicon photomultiplier detectors, digitized at 200 MHz, to record the fast optical flashes of direct Cherenkov emission from EASs. The fluorescence camera (FC) is designed to record tracks of ultraviolet light produced by the development of EASs as viewed from the side. The FC uses multi-anode photomultipler modules digitized at 1 MHz. Below the telescope are two antennas that are part of the EAS radio detector system. 

The tilting mechanism for the PBR telescope is motivated by key scientific and practical requirements. For measurements of EASs via air fluorescence, the energy threshold is lowest and the event rates are highest, with the FC pointing at or near nadir (straight down). The corresponding elevation angle for EAS measurements via direct Cherenkov emission with the CC is just above Earth's limb, about 5.8 degrees below horizontal as seen from the 33~km altitude of the balloon. These Cherenkov emission measurements also provide an in situ demonstration that the CC is functioning properly. The best neutrino sensitivities for the CC are reached by viewing angles a few degrees below the limb. Direct measurements of stars to provide an in situ test of the telescope focus are possible by pointing the FC above Earth's limb.  Finally, on a practical note, the 1364~kg (3000~lb) weight limit allocated to "science" on the 18 MCF super-pressure balloons flown from W\=anaka limits the size and indirectly the field of view of the PBR telescope. These observations and measurements can only be made by tilting the telescope to different angles during flight.

\section{Tilt System Specifications and Design}
\begin{figure}[!h]
\centering 
\includegraphics[width=0.9\textwidth]{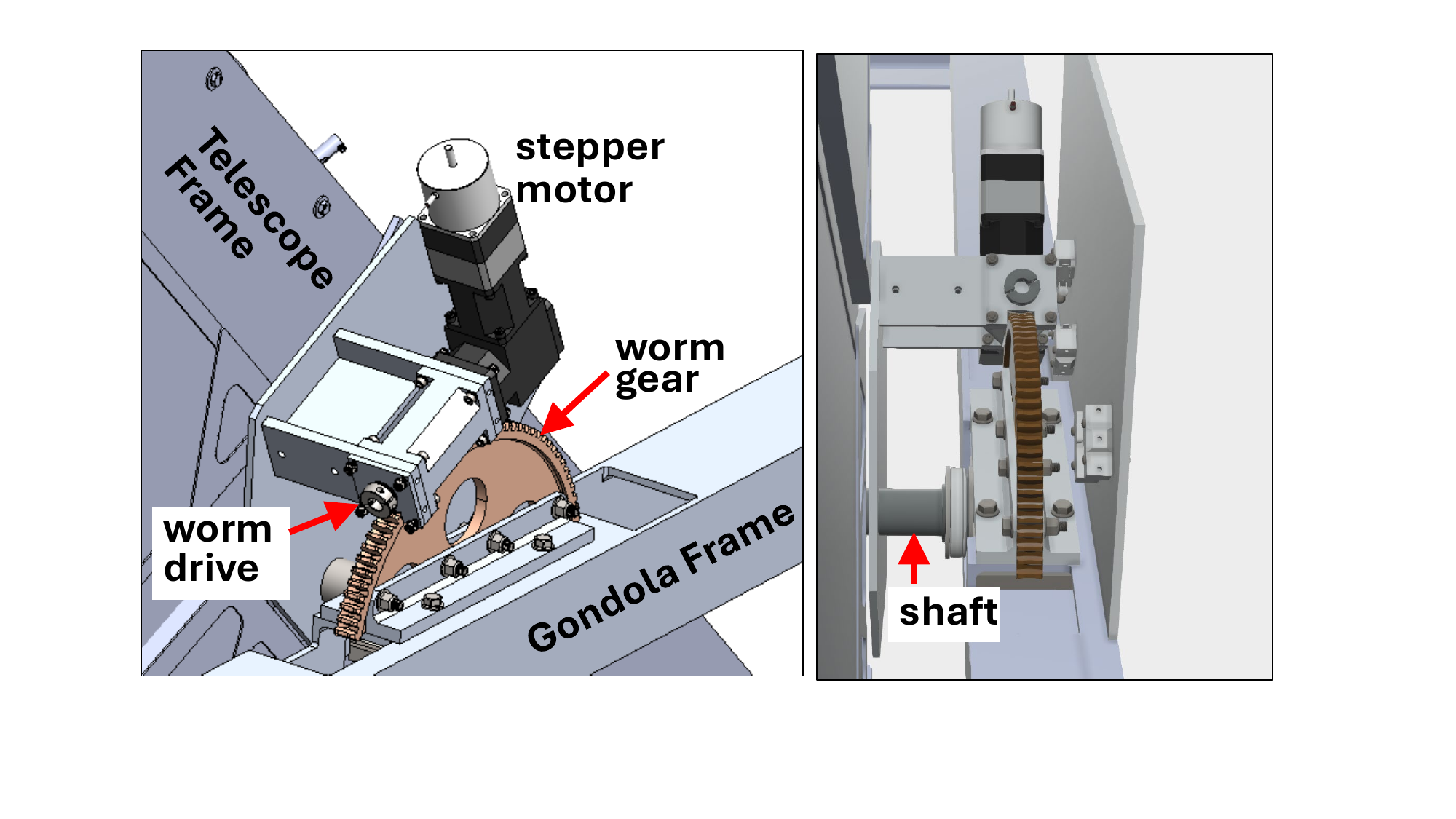}
\caption{The PBR telescope tilt mechanism uses a worm drive turned by a stepper motor to rotate the telescope about a worm gear connected to the gondola frame. Two short shafts, one on each side of the telescope, connect the telescope to the gondola. Each shaft rotates in a custom bearing assembly connected to the gondola. There is one tilt mechanism}
\label{fig:PBR-Tilt-Mechanism}
\end{figure}

The system specifications (Table \ref{tab:spec}) developed for the tilt mechanism (Figure \ref{fig:PBR-Tilt-Mechanism}) meet the scientific and operational requirements. 

\begin{table}[htp]
\centering
\begin{tabular}{| l l l l|}
\hline
Item & Specification & Notes&\\
\hline
Tilt Range & +15\textdegree ~to -90\textdegree & relative to horizontal at balloon&\\
Temperature Range & +30C to -60C &&\\
Pressure Range & sea level to 690 PA (33 km) & 7 mbar, (110,000 ft) &\\
Tilt angle measurement & $\pm$~0.1\textdegree & mechanical axis&\\
Telescope Balance & center of mass +/- 25 mm & About tilt rotation axis &\\
\hline
Select Components &&&\\
\hline
Motor type & stepper & NEMA 23&\\
Controller & programmable logic controller &Siemens 6AG12141AG402XB0 &\\
Tilt Sensor &gravity referenced &Rieker H6MM 2-axis &\\
\hline
\end{tabular}
\caption{PBR Tilt system specifications}
\label{tab:spec}
\end{table}

The telescope is connected to the gondola frame by two short 1.5" stainless steel shafts. One shaft is fixed to each side of the telescope. The other ends of these shafts rotate in custom sleeve bearing assemblies that are bolted to the gondola frame. (The option of using one long single shaft that spanned the telescope was not selected to save weight and to avoid optical and mechanical interference inside the telescope.) The short shafts and bearings are specified to meet the 10g NASA load requirement.  The tilt mechanism uses a worm drive design. A section of a wormgear is bolted to the gondola frame such that its circumference is centered on the 1.5" shaft. A subassembly consisting of a stepper motor driving a worm through a gear reducer is bolted to the telescope frame. As the stepper motor turns, it pushes or pulls the worm against the worm gear to rotate the telescope. The motor subassembly moves with the telescope. 

Attached to this subassembly are two sets of machine tooling balls positioned to trip a pair of switches at either end of the telescope rotation range. Two of the switches are used as limit reference switches. The other two serve as emergency hard stop switches to prevent mechanical collisions between the motor subassembly and the gondola frame if something goes wrong. These switches should never be tripped.

To measure the telescope tilt angle we will fly two identical tilt sensor assemblies, both mounted on the telescope with 0\textdegree calibrated relative to the telescope optical axis. These OTS gravity-based tilt sensors use an internal pendulum mechanism. The accuracy of the tilt angle of the telescope during flight involves several considerations. One question is how accurately can the telescope be tilted to a specified angle.  One step of the stepper motor will rotate the telescope by less than 0.001\textdegree, which will provide very precise tilting of the telescope relative to the gondola. The angle of rotation can be calibrated on the ground using a digital level in terms of motor steps relative to a limit switch. Furthermore, the number of motor steps to travel between the upper and lower limit switches can also be counted on the ground and could occasionally be compared with the values counted in flight to check consistency. A critical question is how accurately can the absolute tilt angle be measured during flight. The absolute accuracy of the tilt sensor is specified as $\pm$~0.1\textdegree. An additional uncertainty is to what extent a small swing of the gondola during flight would affect the pendulum mechanism. Evaluating ways to estimate this is in progress and the overall error budget is under evaluation. For reference, the pixels on the PBR CC and FF have fields of view of 0.2\textdegree~$\times$~0.2\textdegree.

\begin{figure}[!h]
\centering
\includegraphics[width=0.9\textwidth]{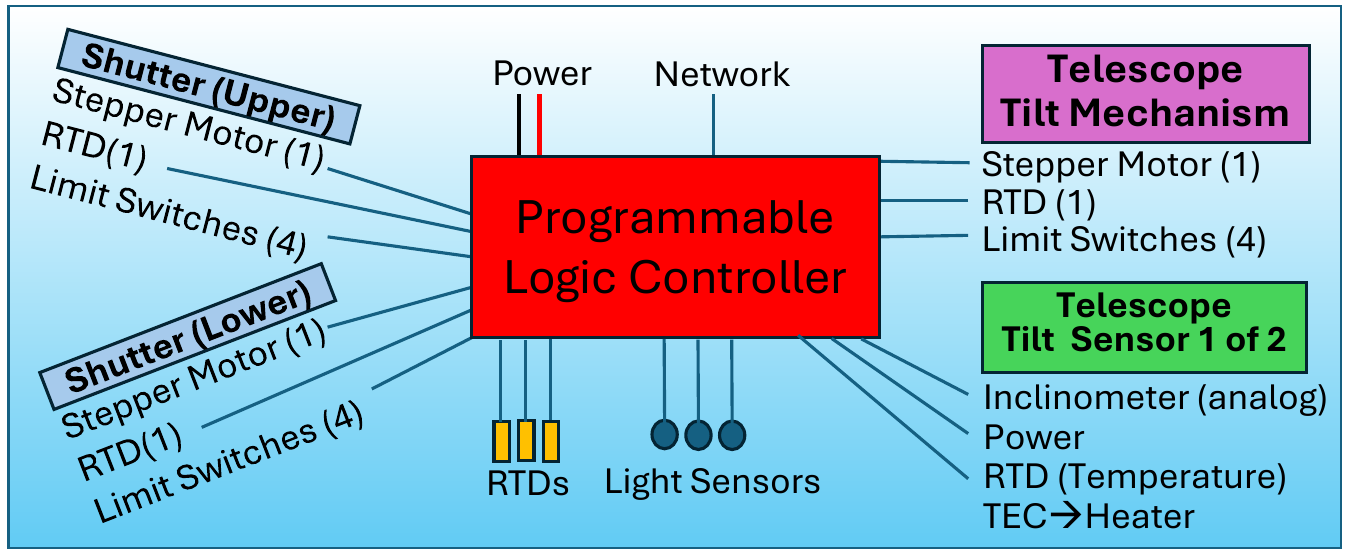}
\caption{Control system architecture of the PBR tilting and telescope shutter systems.}
\label{fig:PBR-Arch}
\end{figure}

The system architecture, shown in Figure \ref{fig:PBR-Arch}, features a programmable logic controller (PLC). Examples of critical systems controlled by PLCs include traffic lights and elevators. PBR uses the same PLC model that was used to control the tilt system and telescope shutters in the EUSO-SPB2 mission. Based on that experience, two PLC IO modules were added to the PBR system to read out resistive temperature devices (RTDs) that measure temperature and to read out additional limit switches. The PLC program used for EUSO-SPB2 is being adapted for PBR to support an expanded command set and return an expanded status packet that contains tilt sensor readings, temperature readings, motor step counts, limit switch statuses, and light sensor readings. 

The tilt sensors and the PLC controller will be maintained at or above a minimum temperature, likely 10~C, using resistive heaters controlled by thermoelectric controllers (TEC). The TEC smoothly ramps the voltage up and down to reduce EMI emission relative to thermal "bang-bang" switches. 
\newpage

\section{Preparation for ground tests}

\begin{figure}[!h]
\centering
\includegraphics[width=0.9\textwidth]{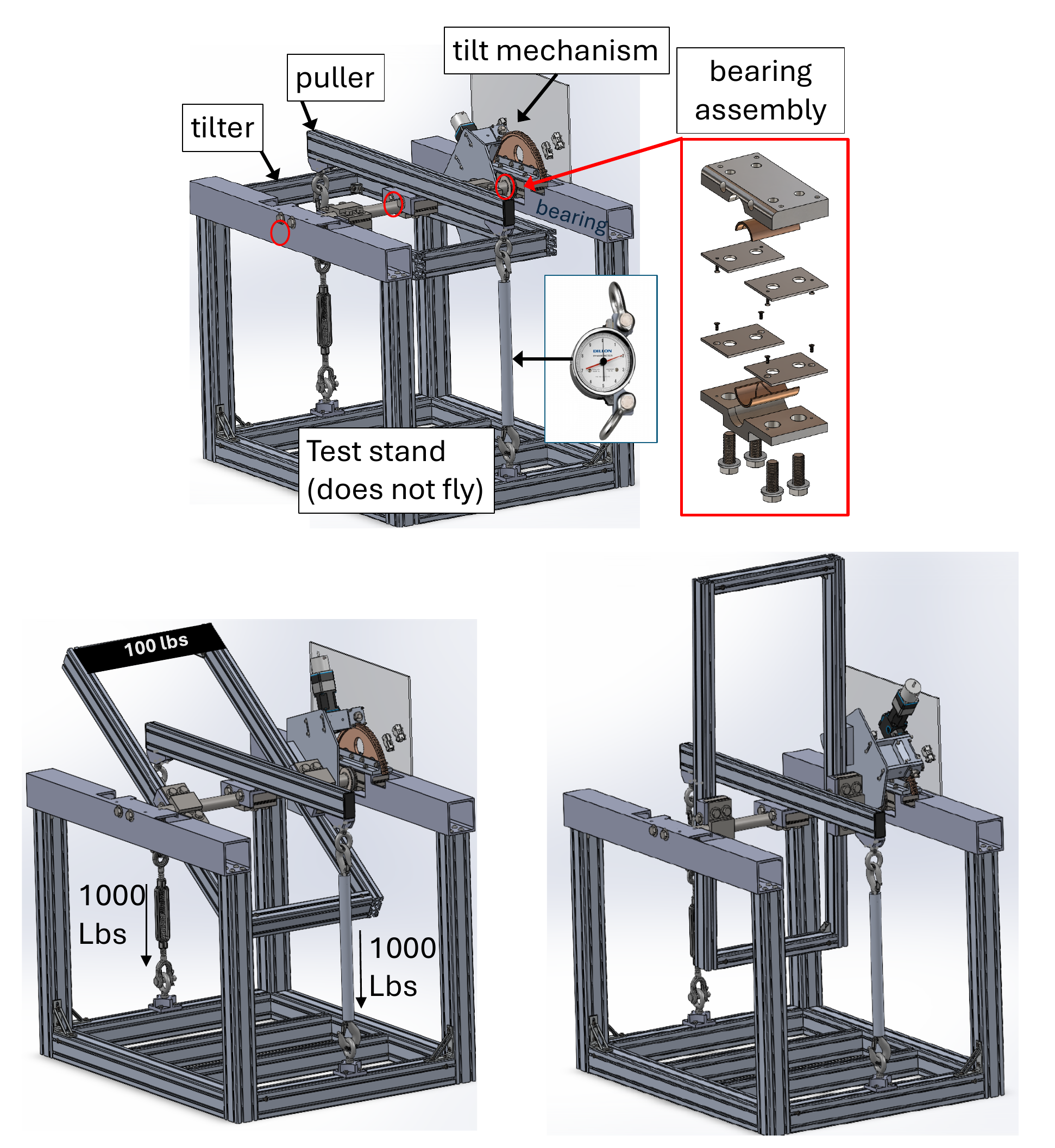}
\caption{To test the PBR tilt system, including the tilt mechanism, bearing assembly, tilt sensors and control system, at 600 PA and -60C and under the expected mechanical load of the 900 kg (2000 Lb.) telescope, a custom test stand is under construction. The stand allows the range of rotation expected for flight, and is designed to fit in the “BEMCO chamber at NASA’s CSBF facility.}
\label{fig:PBR-Tilt-Stamd}
\end{figure}

A custom test stand (figure \ref{fig:PBR-Tilt-Stamd}) is being constructed that will test the tilt mechanism in a large thermal vacuum chamber at the Columbia Balloon Facility. The test stand is arranged to apply the estimated 900 kg (2000 lb) mass of the telescope to the sleeve bearing assemblies while the mechanism rotates. A simulated torque of 22,000 N cm (2000 in lbs) is also applied to simulate the torque arising from a 2.5 cm offset between the rotation shaft axis and the center of mass of the telescope. The tilt sensor assemblies and the PLC control system will also be mounted on the stand for these tests. A critical test will be to find out if the tilt assembly rotates smoothly under these loads at low pressure and -60C and if the tilt sensors measure the tilt angle within or close to the specified accuracy. 

Following the productive desert field test experience of the EUSO-SPB1 and EUSO-SPB2 telescopes, the PBR telescope will also be tested in the desert. The plan is to perform these tests with PBR payload inverted, ie the telescope attached to the gondola and both upside down relative to the flight orientation. The tilt mechanism will provide up-and-down tilting of the telescope.  The gondola will be placed on a rotating turntable driven by a computer controlled slew drive. The turn table assembly will be mounted to the same above-deck trailer used for the EUSO-SPB2 telescopes. This arrangement will allow the FC and CC to be scanned vertically and horizontally across various light sources including a laser and point light sources, for example. 

\section{Conclusion}
The fabrication of the tilt mechanism and the tilt sensor assemblies is making good progress. The design is based on the experience of the EUSO-SPB2 design and mission. The PBR tilt system will enable the collection of a broad range of astroparticle and instrument characterization data, and pioneer the evolution of ground-based methods to measure cosmic rays and neutrinos from the vantage point of near space as a precursor to a space-based mission such as POEMMA.

\vskip 0.2in
\noindent{\bf Acknowledgements}
\vskip 0.2in

Work supported by NASA awards 11-APRA-0058, 16-APROBES16-0023, 17-APRA17-0066, NNX17AJ82G, NNX13AH54G, 80NSSC18K0246, 80NSSC18K0473, 80NSSC19K0626, \hfil \break 
80NSSC18K0464, 80NSSC22K1488, 80NSSC19K0627 and 80NSSC22K0426, the French space agency CNES, National Science Centre in Poland grant n. 2017/27/B/ST9/02162, and by ASI-INFN agreement n. 2021-8-HH.0 and its amendments. This research used resources of the US National Energy Research Scientific Computing Center (NERSC), the DOE Science User Facility operated under Contract No. DE-AC02-05CH11231. We acknowledge the NASA BPO and CSBF staffs for their extensive support. We also acknowledge the invaluable contributions of the administrative and technical staffs at our home institutions.

\bibliography{icrc2025_bib}

\providecommand{\href}[2]{#2}\begingroup\raggedright\begin{thebibliography}{10}

\bibitem{ICRC2025Eser}
J.~Eser, A.V.~Olinto, G.~Osteria et~al. ~{\emph{PoS} {\bfseries ICRC2025} (2025) 934}.

\bibitem{FT-paper}
J.H.~Adams, Jr. et~al. ~\href{https://doi.org/10.1016/j.astropartphys.2024.103046}{\emph{Astropart. Phys.} {\bfseries 165} (2025) 103046} [\href{https://arxiv.org/abs/2406.13673}{{\ttfamily 2406.13673}}].

\bibitem{Adams:2025owi}
J.H.~Adams et~al. ~ \href{https://arxiv.org/abs/2505.20762}{{\ttfamily 2505.20762}}.

\bibitem{POEMMA:2020ykm}
{\scshape POEMMA} collaboration ~\href{https://doi.org/10.1088/1475-7516/2021/06/007}{\emph{JCAP} {\bfseries 06} (2021) 007} [\href{https://arxiv.org/abs/2012.07945}{{\ttfamily 2012.07945}}].

\bibitem{EASCherSim}
A.~Cummings, R.~Aloisio, J.~Eser and J.~Krizmanic ~\href{https://doi.org/10.1103/PhysRevD.104.063029}{\emph{Phys. Rev. D} {\bfseries 104} (2021) 063029} [\href{https://arxiv.org/abs/2105.03255}{{\ttfamily 2105.03255}}].

\bibitem{Reno:2019jtr}
M.H.~Reno, J.F.~Krizmanic and T.M.~Venters ~\href{https://doi.org/10.1103/PhysRevD.100.063010}{\emph{Phys. Rev. D} {\bfseries 100} (2019) 063010} [\href{https://arxiv.org/abs/1902.11287}{{\ttfamily 1902.11287}}].

\bibitem{Venters:2019xwi}
T.M.~Venters et~al. ~\href{https://doi.org/10.1103/PhysRevD.102.123013}{\emph{Phys. Rev. D} {\bfseries 102} (2020) 123013} [\href{https://arxiv.org/abs/1906.07209}{{\ttfamily 1906.07209}}].

\bibitem{Venters_2021ICRC}
T.M.~Venters, M.H.~Reno and J.F.~Krizmanic ~{\emph{PoS} {\bfseries ICRC2021} (2021) 977}.

\bibitem{Eric:ICRC}
E.~Mayotte et~al. ~{\emph{PoS} {\bfseries ICRC2025} (2025) 541}.

\bibitem{Francesco:ICRC}
F.~Cafagna et~al. ~{\emph{PoS} {\bfseries ICRC2025} (2025) 483}.

\bibitem{Valentina:ICRC}
V.~Scotti et~al. ~{\emph{PoS} {\bfseries ICRC2025} (2025) 997}.

\end{thebibliography}\endgroup

    \newpage
{\Large\bf Full Authors list: The JEM-EUSO Collaboration}

\begin{sloppypar}
{\small \noindent
M.~Abdullahi$^{ep,er}$              
M.~Abrate$^{ek,el}$,                
J.H.~Adams Jr.$^{ld}$,              
D.~Allard$^{cb}$,                   
P.~Alldredge$^{ld}$,                
R.~Aloisio$^{ep,er}$,               
R.~Ammendola$^{ei}$,                
A.~Anastasio$^{ef}$,                
L.~Anchordoqui$^{le}$,              
V.~Andreoli$^{ek,el}$
% ...
}
\end{sloppypar}
\vspace*{.3cm}

{ \footnotesize
\noindent
%
% ...
$^{cb}$ Universit\'e de Paris, AstroParticule et Cosmologie (CNRS), Paris, France\\
% ...
$^{ef}$ Istituto Nazionale di Fisica Nucleare (INFN), Sezione di Napoli, Naples, Italy\\
% ...
$^{ei}$ Istituto Nazionale di Fisica Nucleare (INFN), Sezione di Roma Tor Vergata, Rome, Italy\\
% ...
$^{ek}$ Istituto Nazionale di Fisica Nucleare (INFN), Sezione di Torino, Turin, Italy\\
$^{el}$ Universit\`a di Torino, Dipartimento di Fisica, Turin, Italy\\
% ...
$^{ep}$ Gran Sasso Science Institute (GSSI), L’Aquila, Italy\\
$^{er}$ Istituto Nazionale di Fisica Nucleare (INFN), Laboratori Nazionali del Gran Sasso, Assergi, Italy\\
% ...
$^{ld}$ University of Alabama in Huntsville, Huntsville, AL, USA\\
$^{le}$ City University of New York (CUNY), Lehman College, Bronx, NY, USA\\
% ...
}

\end{document}